# Calculation of High-current Linear Proton Accelerator for the Energy of 80 MeV

*S. N. Dolya and I. Sarhadov*

*Joint Institute for Nuclear Research, Joliot - Curie 6, Dubna, Russia, 141980*

**Abstract**

The article considers an opportunity of pulsed acceleration of the proton beam with current $I_p = 0.7$ A in a spiral waveguide. The accelerator consists of three parts. The energy of the proton injection is 50 keV. In the first part of the accelerator having length $L_{acc1} \approx 1.4$ m, the protons are accelerated to the energy of $Є_{1fin} = 0.8$ MeV. Consumed high-frequency power of this part of the accelerator is equal to $P_1 = 0.8$ MW. In the second part of the accelerator having the length $L_{acc2} \approx 2$ m, protons are accelerated till the energy of $Є_{2fin} = 5$ MeV. The consumed high-frequency power in the second part of the accelerator is $P_2 = 4$ MW. The third part of the accelerator consists of 8 sections, each 7 m long. The radial focus of the proton beam in the accelerator is carried out by means of the magnetic field of 10 T, generated by a superconducting solenoid.

## 1. Introduction

In Dubna there is an operating pulsed reactor IBR - 2 [1], whose complex of buildings includes a facility prepared for locating an accelerator 200 m long. As [2] proposed this facility was assumed to accommodate a high-current proton accelerator with beam current $I_b = 0.7$ A and beam energy $Є = 1$ GeV. Irradiating the subcritical nuclear reactor with the beam of this accelerator it will be possible to obtain a neutron flux by several times exceeding the flow of neutrons from the pulsed neutron source ESS [3].

The main accelerator irradiating the subcritical nuclear reactor [2] is assumed to be superconducting. The injector for this superconducting accelerator should to be a proton accelerator [4] with a proton beam energy of $Є = 80$ MeV. The acceleration of protons till the energy $Є = 80$ MeV is suggested to carry out in a linear accelerator based on the spiral waveguide.

The spiral waveguide is a conventional waveguide coaxial cable whose central conductor is wound as a spiral over a dielectric frame, Fig. 1.

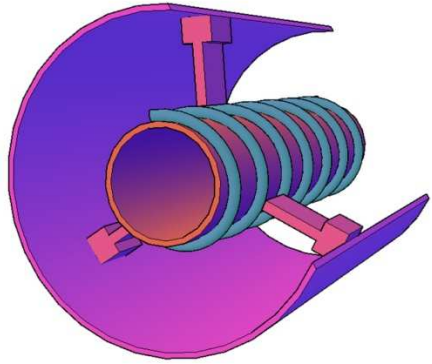

Fig. 1. Scheme of the spiral waveguide.

This spiral structure may have propagation of slow electromagnetic waves which can be used to accelerate protons. A detailed theoretical analysis of proton acceleration with the initial energy of 1 MeV to the final energy of 10 MeV was performed in 1959, [5]. This article, however, did not consider either the beam focusing in the transverse direction or the RF power transfer into the beam.

Below there is a review of the dynamics of protons in the linear accelerator built on the base of the spiral waveguide. The phase motion of the protons will be taken into account as well as the effect of the Coulomb repulsion of particles in the accelerator.

Fig. 2 shows a scheme of the accelerator.

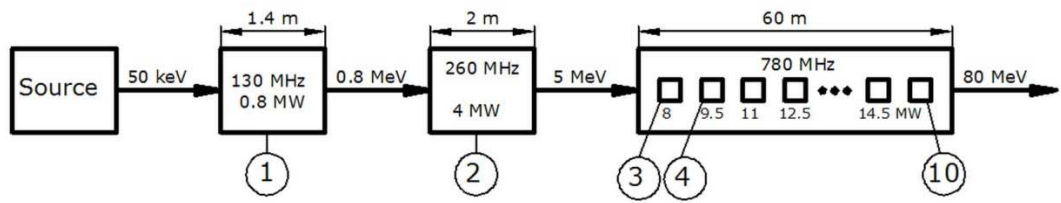

Fig. 2. The circuit of the accelerator.

**2. Longitudinal motion of particles in the beam**

Any spiral waveguide is a structure where the acceleration occurs in the traveling wave, which like in [6] has been chosen as a $\cos(k_3 z - \omega t + \varphi_s)$. For a synchronous particle its velocity $v_p = z/t$ always coincides with the phase velocity of the wave $v_{ph} = \omega/k_3$ so that the synchronous particle is always in the same phase of the wave, i.e. in the electric field $E = E_0 * \cos\varphi_s$. When the synchronous particle is accelerated its velocity continuously grows up according to the following equation:

$$mc^2 d\gamma/dz = eE_0\cos\varphi_s, \qquad (1)$$

where $m = 1.67264 * 10^{-24}$ g – the rest mass of the proton, with $c = 2.99792*10^{10}$ cm/s is the velocity of light in vacuum, e - elementary charge, $\gamma = (1-\beta^2)^{-1/2}$ - relativistic factor, $\beta = v_p/c$ - the velocity of protons, expressed in terms of the light velocity.

Continuous acceleration of protons by the wave means that in the slowing down structure, i.e., in the spiral waveguide, it is necessary to continuously accelerate the wave. For the spiral waveguide where the spiral is wound over the cylindrical frame, it is possible to obtain the increase of the wave velocity by increasing the winding step of the spiral. The dispersion equation was obtained in [7]. It relates the phase velocity of the wave in the structure and the spiral waveguide parameters: spiral winding radius $r_0$, the radius R of the screen and a spiral winding step.

The role of the external screen is primarily to reduce the tension of the electric field on the axis of the spiral, which can be very significant when the screen is close to the spiral. [8]

Fig. 3 shows the dependence of decreasing the amplitude of the electric field on the axis of the spiral on the radius of the screen.

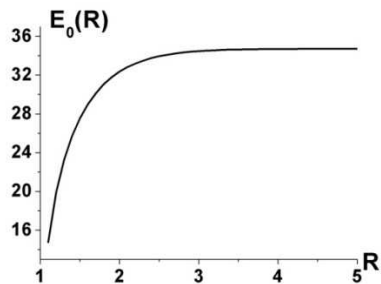

Fig. 3 The dependence of the tension amplitude of the electric field on the axis of the spiral $E_0$ (kV/cm) on the screen radius R (cm).

Fig. 3 shows that for the selected [8] screen of radius R = 3 cm there is almost no electric field reducing on the spiral axis. The screen is located far away from the spiral, and does not influence the field structure in the spiral.

Let us choose the voltage at which the ion source is located relatively the ground to be equal to: $U_{es}$ = 50 kV. The protons with this energy have the velocity, expressed in terms of the light velocity equal to β = 0.01, the dimensional velocity in this case is equal to $v_{inj}$ = 3*10$^8$ cm/s.

*2. 1. The Coulomb repulsion in the longitudinal direction*

At a large beam current, i.e., at a large number of particles in bunches, it may happen that the focus force affecting on the bunch from the wave will not be enough to resist this repulsion.

The frequency square $\Omega^2_{ph}$ of phase oscillations is connected with the back force acting on the protons from the wave by the following relationship: F '= $m\Omega^2_{ph}z'$, where F 'is the force acting from the wave on the protons. This force has explicitly several multiply components: coordinate z', and m denotes the mass of a proton.

To find the Coulomb repulsion force, it is necessary to make some assumptions about the shape of the bunch, and the distribution of particles in it. We assume that the bunch has a form of an ellipsoid with semi axis's: $r_b$ and Δz ', and the charge distribution in the ellipsoid, we assume to be uniform. The volume of the ellipsoid is equal to the following:

$$V_b = (4/3)\pi r^2_b \Delta z'. \qquad (2)$$

Let the total number of protons in the bunch is equal to N. This number of protons is related to the proton beam current by the following ratio:

$$I_b = eN * f_1, \qquad (3)$$

where: e = 1.6*10$^{-19}$ C - elementary charge, $f_1$ = 1.3*10$^8$ Hz – is the accelerating frequency.

Let us find the number N of protons in the bunch for the proton beam current $I_b$ = 0.7 A. Substituting the figures into the formula (3), we find that N = 3.3*10$^{10}$ protons per bunch. Now, if we divide this number by the volume of the proton bunch, we find n - the volume density of protons in the bunch. We assume that the radius of the beam is equal to: $r_b = 0.5$ cm. This radius, from one hand, should be smaller than the spiral radius $r_0 = 1$ cm, from the other hand, at too small radius of the beam the volume proton density n in the bunch will be difficult to hold by means of the high-frequency wave field.

Relative to the longitudinal semi size of the bunch $\Delta z'$ we can say the following: the phase region occupied by the bunch on the phase plane is equal to $3\varphi_s$ and that is why the value $2*\Delta z'$ may be defined as follows:

$$2\Delta z' = 3\varphi_s * \lambda_0 \beta / 2\pi, \qquad (4)$$

where: $\lambda_0 = c/f_1 = 230$ cm, i.e., the wavelength in free space in the first section.

From the above it is clear that the shortest length of the bunch will be at the beginning of the acceleration after grouping, where $\beta_{in} = 0.01$. From (4) we can find the value of $\Delta z'$- longitudinal semi-size of the beam. In the first section we choose the synchronous phase value $\varphi_s = 80^0$, $\cos\varphi_s = 0.17$. Then we find that in this case for the selected parameters, $\Delta z' = 3\varphi_s * \beta\lambda_0/4\pi = 0.77$ cm.

In this accelerator the Coulomb repulsion is the most important process, as it is limiting the beam current. Consider this option in detail.

Let the bunch has an elliptical shape with semi axis's $r_b$ and $\Delta z'$. The electric field inside the bunch can be written as follows in [9]:

$$E_C = 4\pi e n \xi z', \qquad (5)$$

where z'- longitudinal coordinate counted from the center of the bunch, n- density of protons in the bunch, $\xi$ - coefficient of the bunch form.

The proton density can be calculated if the total number of protons in the bunch $N = 3.3*10^{10}$ divided by its volume $V_b = (4/3) \pi r^2_b \Delta z' = 0.77$ cm$^3$. Having fulfilled this calculation, we find that the density of protons in bunch is equal to:

$n = 4.3*10^{10}$p/cm$^3$. The shape coefficient of the bunch can be found according to the formula given in [9]:

$$\xi = (1-l^2_1)*\{\ln[(1+l_1)/(1-l_1)] -2l_1\}/l^3_1, \qquad (6)$$

where $l_1 = (1-r^2_b/\Delta z'^2)^{1/2}$. Substituting numbers in this expression, we find that the eccentricity of the bunch in this case at the beginning of the acceleration in the first section of the spiral waveguide is equal to: $l_1 = 0.76$.

The shape coefficient of the bunch $\xi$, according to formula (6) is equal to: $\xi = 0.25$.

Substituting this coefficient in formula (5) we obtain an expression for the Coulomb field acting on the proton in the longitudinal direction in the bunch:

$$E_C = \pi enz'. \qquad (7)$$

And, finally, we find the square of the Coulomb frequency in the bunch:

$$\omega^2_C = \pi e^2 n/m. \qquad (8)$$

This square of the Coulomb frequency should be compared with the square of the frequency of the longitudinal phase oscillations.

We estimate the Coulomb field for the very beginning of the acceleration of a bunches located in the first section of the beam after grouping. Calculating the square of the frequency of the phase oscillations of protons in bunches at the beginning of the first section of the accelerator we obtain the following:

$$\Omega^2_{ph} = \omega^2(eE_0\lambda_0\sin\varphi_s/2\pi mc^2\beta_s) = 6.6*10^{17}*0.073 = 4.9*10^{16}.$$

For the same region of the accelerator we calculate the frequency of oscillations $\omega_C^2$ and find:

$$\omega_C^2 = \pi e^2 n/m = 3.14*25*10^{-20}*4.3*10^{10}/1.7*10^{-24} = 1.9*10^{16}.$$

We see that the square of the Coulomb frequency is less than the square of the phase oscillation frequency. It proves that in this case the phase oscillations will be stable and Coulomb repulsion of particles in bunches will only result in reducing of the frequency of phase oscillations.

The ratio of the squares of the frequencies, i.e. the ratio of the square of the frequency of the phase oscillations $\Omega^2_{ph}$ to the square of the frequency of Coulomb repulsion $\omega_C^2$ will change during the acceleration. This is caused by the fact that the longitudinal size of the bunch $\Delta z'$ increases while accelerating: $\Delta z' = (3/4\pi)\varphi_s\beta\lambda_0$. The frequency of phase oscillations decreases while $\beta$ is growing. The ratio of the frequency squares is equal to the following

$$\Omega^2_{ph}/\omega_C^2 = (2/3)\omega^2 E_0 \lambda_0 \sin\varphi_s * r^2_b \Delta z'/(\beta c^2 * \pi eN). \qquad (9)$$

Equation (9) can be rewritten as:

$$\Omega^2_{ph}/\omega_C^2 = \omega^2 E_0 \lambda_0 r^2_b \varphi_s \sin\varphi_s / 2\pi^2 ec^2 N. \qquad (10)$$

Tension $E_0$ of the wave field in the considered case reduces from the beginning to the end of acceleration. This can result to the following: the Coulomb repulsion force will become greater than the focusing power of the wave.

### 3. The required high-frequency power

The logic of the calculation of the accelerating section load by the beam is the following. The power $P_b = \int I_b * E_0 * \cos\varphi_s dz$ is transmitted into the beam. It means that it is necessary to increase the magnitude of the introduced power in comparison with the power calculated according to the formula for the flux.

In the first part of the accelerator, the electric field $E_0 = 0.2$ MV/m, $\cos\varphi_s = 0.17$. In this section the protons obtain the energy $Є = 800$ keV, at the beam current $I_b = 0.7$ A. Thus the transmitted beam power $P_b = 0.8$ MeV * 0.7A = 560 kW. This means that to the power $P = 200$ kW, which goes to form the magnetic field, it is necessary to add $P_b = 0.8$ MeV*0.7 A = 560 kW, which will be transferred to the beam. As a result, the required total power, introduced into the first part of the accelerator should be equal to $P_1 = 800$ kW.

Fig. 4 shows the corresponding graphs taking into account the beam influence on high frequency power.

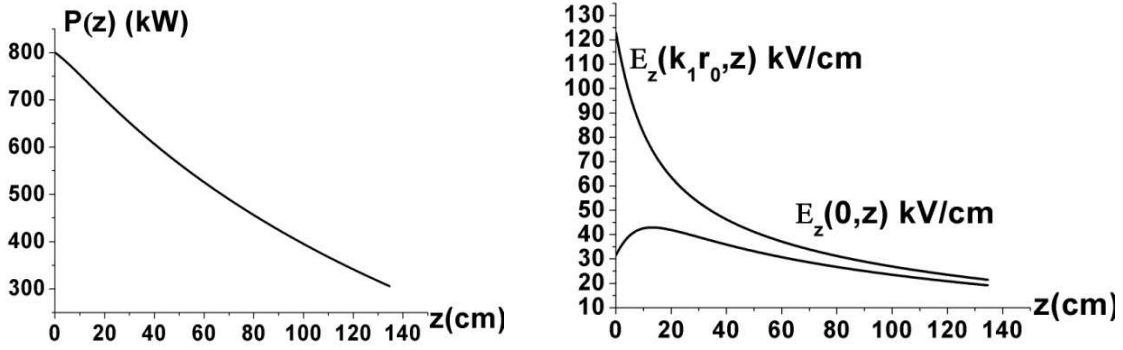

Fig. 4. The dependences of the power and electric field tensions on the acceleration length.

**4. The radial motion of protons in the beam**

The wave field at radius $r_b = 0.5$ cm can be found as $E_r = -r_b/2 \cdot dE_z/dz = (\pi r_b/\beta\lambda_0)E_0$. This is a large value. Under the influence of this field the protons are accelerated in a radial direction, and can go far enough away from the beam axis, that can result in increasing the radial size of the beam. As a result, the beam with a large radial dimension may hurt the spiral. It should be emphasized that the radial electric field of the wave increases with increasing the radius. In this case the situation becomes even more complicated.

Fig. 5 shows the dependence of the radius deviation of the proton being in the initial moment at the distance $r_b = 0.5$ cm from the beam axis. The proton is in the synchronous phase of the wave. The proton deviation is caused by the influence of the radial field of the wave and Coulomb repulsion calculated by the model of uniformly charged ellipsoid [9]:

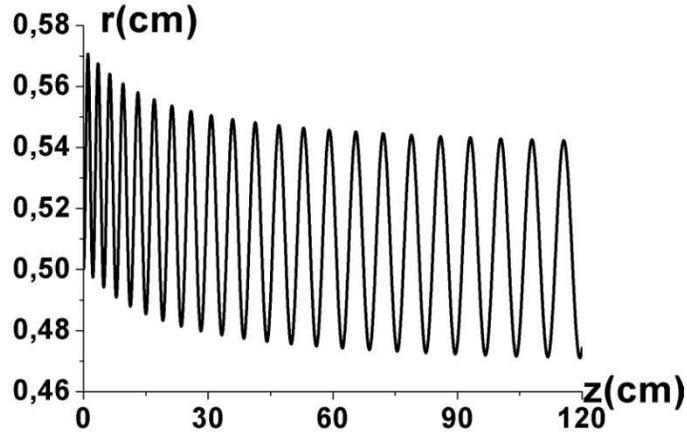

Fig. 5. Dependence of radial deviation of the proton-axis located in the initial time at radius $r_b = 0.5$ cm in the synchronous phase of the wave field.

It is seen that the maximum additional deviation of the proton from the accelerator axis is of the order of 1 mm.

It can be seen that the length of the acceleration for the proton, located on the border of the beam $r_b = 0.5$ cm is by about 15% smaller than for the proton, accelerated along the axis. This is caused by the rapid growth of the accelerating field along the radial coordinate. To synchronously accelerate the proton located on the axis and the proton located at radius $r_b = 0.5$ cm, it is necessary to have the synchronous phase on the radius $r_b = 0.5$ cm greater than at the axis. It means that the following condition must be fulfilled: $E_{01}*\cos\varphi_{s1} = E_{02}*\cos\varphi_{s2}$, where $E_{01}$ and $E_{02}$ are the amplitudes of the electric field tensions on the acceleration axis and on the radius of the beam boundary, $\varphi_{s1}$, $\varphi_{s2}$ are the corresponding synchronous phases.

In our case, when the amplitude of the accelerating field is by 15% bigger at the boundary of the beam in comparison with the beam axis, the synchronous phase at the beam boundary should be equal to $\varphi_{s2} = 81.5^0$, instead of $\varphi_{s1} = 80^0$ for the beam axis. Note that the horizontal size of the separatrix is proportional to $\varphi_s$. Thus, at this increase of the separatrix size all the particles of the beam located in one transverse cross-section will be captured, each in its separatrix. The dependence of the horizontal sizes of the separatrix on the radial deviation from the axis of the system will not affect the capture of protons in the acceleration mode.

Table 1 summarizes the main parameters of the first part of the accelerator.

Table 1. Parameters of the initial part of the accelerator.

| Option | Value |
| --- | --- |
| The voltage of the proton source is, kV | 50 |
| Bunched beam current, mA | 700 |
| The frequency of the acceleration f, MHz | 130 |
| The tension of the focusing magnetic field, T | 10 |
| The average tension of electric field $E_0$, MV /m | 0.6 |
| The cosine of the synchronous phase, $\cos\varphi_s$ | 0.17 |
| High-frequency power, MW | 0.8 |
| The length of the accelerator, m | 1.4 |

**5. The second part of the accelerator**

We consider the motion of a synchronous particle in the accelerator which is assumed to consist of three parts. In the first part of the accelerator the protons must be accelerated from the energy equal to $\mathcal{E}_{in1} = 50$ keV till the energy $\mathcal{E}_{fin1} = 800$ keV, that corresponds to the velocity of the proton, approximately equal to $\beta_{fin1} = 0.04$. In the second part of the accelerator, protons must be accelerated till the energy $\mathcal{E}_{fin2} = 5$ MeV, that corresponds to the velocity of protons $\beta_{fin1} = 0.1$. For the third part of the accelerator the injection energy will be equal to the final energy achieved in the second part of the accelerator, i.e. $\mathcal{E}_{in3} = \mathcal{E}_{fin2} = 5$ MeV, and the final energy must be equal to: $\mathcal{E}_{fin2} = 80$ MeV, that corresponds to the velocity of protons $\beta_{fin3} = 0.4$.

Fig. 6. shows the dependence of the dimensionless velocity increase, β, of the electric field tension, a winding step of the spiral and decrease of the high-frequency power, on the length of the acceleration:

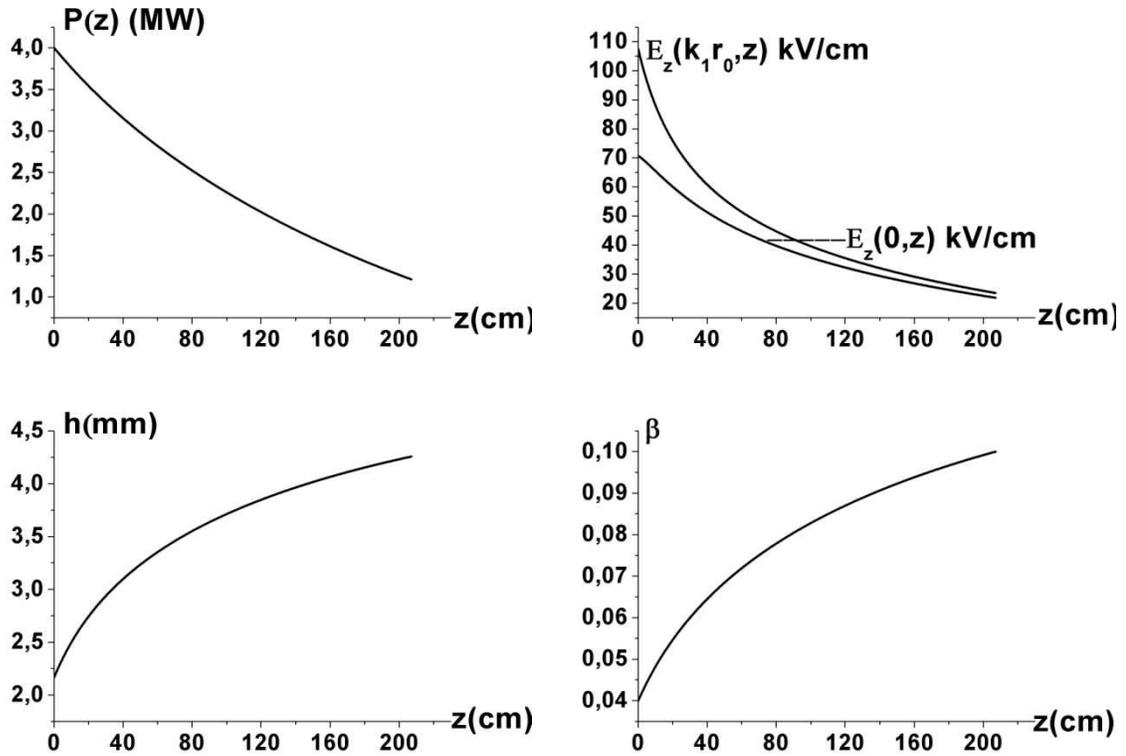

Fig. 6. The dependence of the power, field strength, the winding step of the spiral, and the growth of the relative velocity of the acceleration length. Table 2 shows the main parameters of the second accelerator.

Table 2. Parameters of the second part of the accelerator.

| Option | Value |
|---|---|
| The initial energy, keV | 800 |
| Bunched beam current, A | 0.7 |
| The frequency of the acceleration f, MHz | 260 |
| The tension of the focusing magnetic field, T | 10 |
| The average tension of electric field $E_0$, MV /m | 2.5 |
| The cosine of the synchronous phase, $\cos\varphi_s$ | 0.5 |
| High-frequency power, MW | 4 |

| The length of the accelerator, m | 2 |

## 6. The third part of the accelerator

In the third part of the accelerator the spiral radius remains the same: $r_0 = 1$ cm. We choose a synchronous phase in the third part of the accelerator: $\varphi_{s3} = 60_0$, $\cos\varphi_{s3} = \frac{1}{2}$. The acceleration frequency in the third part of the accelerator we have chosen by 6 times larger than in the first part of the accelerator: $f_3 = 780$ MHz. Figure 7 shows the distributions of the parameters of the third part of the accelerator.

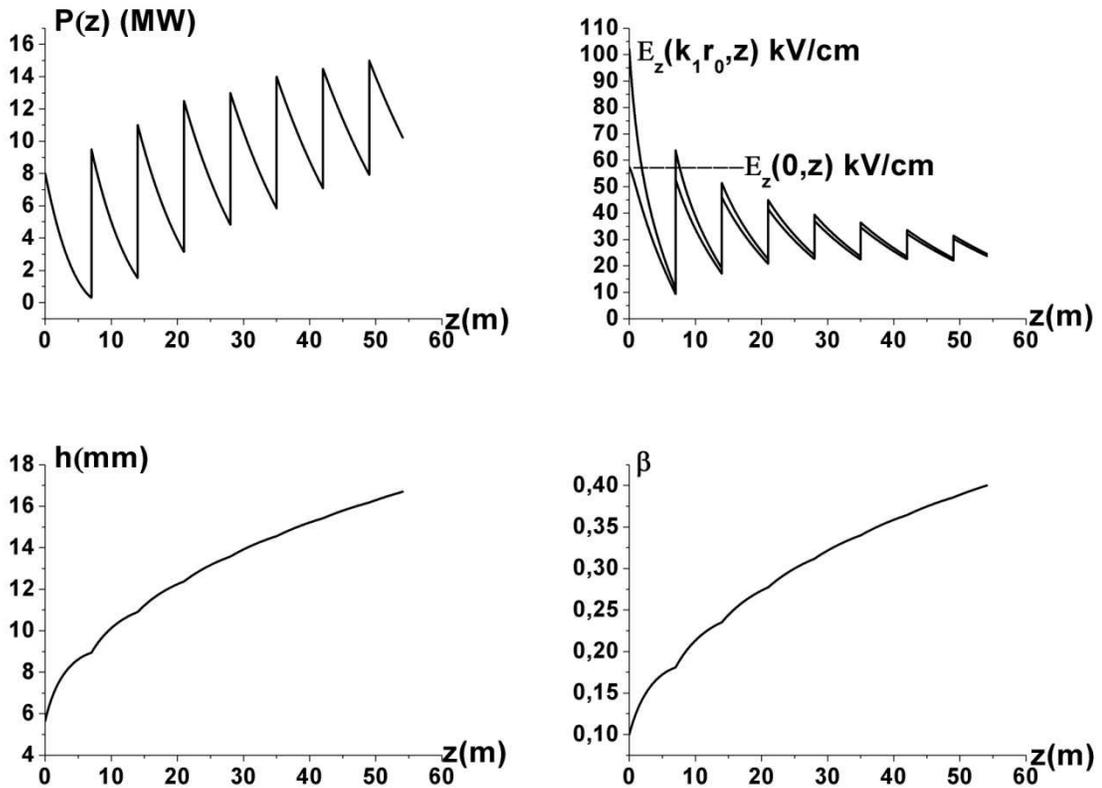

Fig. 7. Distributions of phase velocity, electric field tension, spiral winding step and high-frequency power along the third part of the accelerator.

As a result, after acceleration of protons in 8 sections of the third part of the accelerator, they will obtain energy $\varepsilon = 80$ MeV and their relative velocity will reach the value: $\beta = 0.4$. The length of this part of the accelerator is $L_{acc3} = 7$ m * 8 = 56 m. In this case, in the sections there is still a large high-frequency power, which must be output from the sections to the coordinated load.

Another selection of parameters is also possible. In this case, the lengths of the sections are chosen in such a way that the power in the sections would remain P = 1 MW in the accelerating sections. The corresponding distributions are shown in Fig. 8.

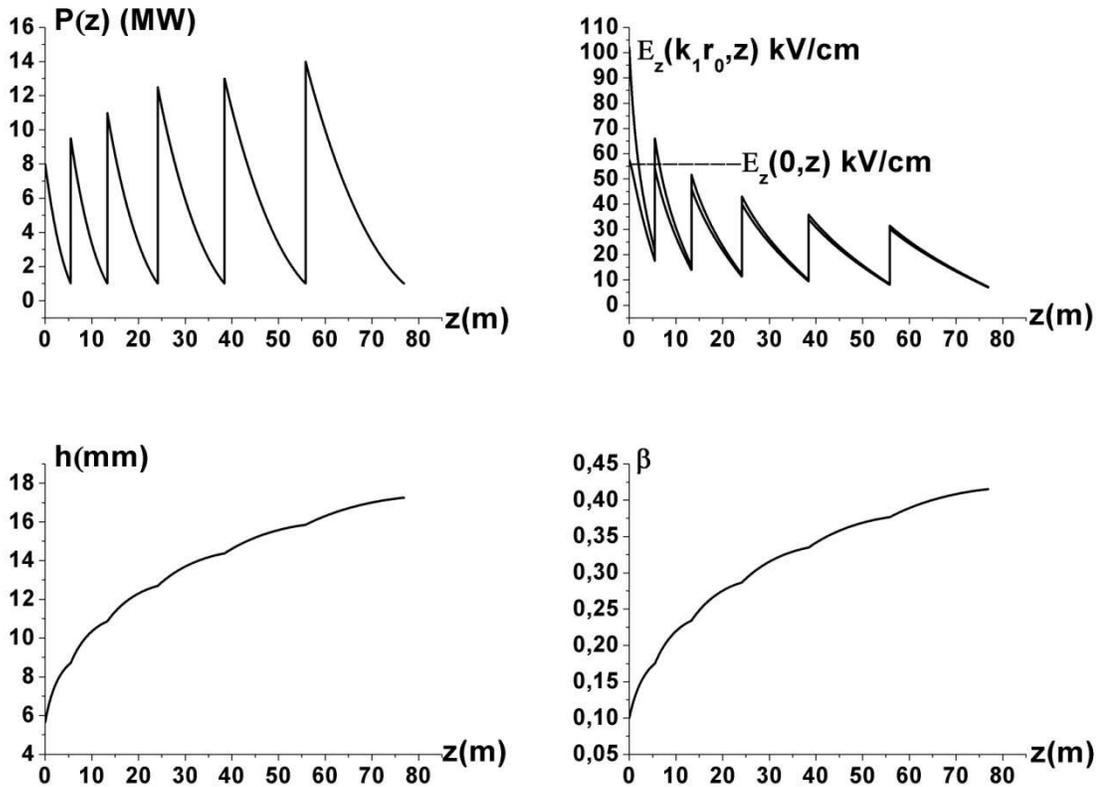

Fig. 8. Dependences of power reduction, distribution of electric field tension, spiral winding step and increase of the proton velocity on the acceleration length.

It is seen that in this case, to achieve an energy $\mathcal{E} = 80$ MeV, it is required to increase the acceleration length till the value of $L_{acc3} = 80$ m.

Table 3 shows the parameters of the third part of the accelerator.

Table 3. Parameters of the third part of the accelerator

| Option | Value |
|---|---|
| Initial energy of the proton, MeV | 5 |
| Bunched beam current, A | 0.7 |
| The frequency of the acceleration f, MHz | 780 |
| The tension of the focusing magnetic field, T | 10 |
| The average tension of electric field $E_0$, MV/m | 3 |
| The cosine of the synchronous phase, $\cos\varphi_s$ | 0.5 |
| High-frequency power, MW | 8, 9.5, …15 |
| The length of the accelerator, m | 60 |

**7. Conclusion**

From the above it can be seen that the acceleration of the proton beam with a current of $I_b = 0.7$A is possible in a spiral waveguide. The accelerator should consist of several parts. The injection energy in the first part of the accelerator is 50 keV and at a length of 1.4 m the protons are accelerated to the energy of 0.8 MeV. In the second part of the accelerator, having a length of 2 m, the protons will be accelerated to the energy of 5 MeV. The third part of the accelerator should consist of eight sections, each 7 m long. After acceleration the protons will reach the energy of 80 MeV. The focusing of the protons in the radial direction should be carried out by a solenoidal magnetic field of H = 10 T.

The impulse current of the proposed accelerator, $I_b = 0.7$ A, significantly exceeds the currents of all the known high-current linear proton accelerators.

References

1. https://ru.wikipedia.org/wiki/ИБР_2

2. S. N. Dolya, A multy beam proton accelerator, http://arxiv.org/ftp/arxiv/papers/1509/1509.04158.pdf

3. https://en.wikipedia.org/wiki/European_Spallation_Source

4. https://www.researchgate.net/publication/287285991_Proton_LINAC_Using_Spiral_Waveguide_with_Finite_Energy_of_80_MeV

5. W. Muller and J. Rembser, Reflections on construction of a linear accelerator for protons, Nuclear instruments and methods, v. 4, iss. 4, p. 202-212, 1959, North-Holland publishing co.

6. I. M. Kapchinsky, Particle dynamics in linear resonance accelerators, Moscow, Atomizdat, 1966

7. A. I. Akhiyezer, Ya. B. Fainberg, Slow down electromagnetic waves, UFN,

v. 44, is. 7, p. 322, 1951, http://ufn.ru/ru/articles/1951/7/a/

8. S. N. Dolya, K.A. Reshetnikova, Linear Accelerator $C^{+6}$ Ions, as Injector for Synchrotron, Intended for Hadrons Therapy,

https://arxiv.org/ftp/arxiv/papers/1307/1307.6302.pdf

9. A. I. Akhyezer, G. Ya. Lubarsky, L. E. Pargamanik, Ya. B. Fainberg,

Preliminary grouping and dynamics of a proton beam in a linear accelerator, In book: Theory and Design of Linear Accelerators, Gosatomizdat, Moscow, 1962

10. D. V. Karetnikov, I. N. Slivkov, V. A. Teplyakov, e. a., Linear accelerators of ions, Gosatomizdat, Moscow, 1962